\documentclass{aa}
\usepackage{times}
\usepackage{amssymb}
\usepackage{graphicx}
\begin{document}

   \thesaurus{03     % A&A Section 3: Extragalactic astronomy 
              (13.07.01; % Gamma rays: bursts,
               11.19.3;  % Galaxies: starburst,
               09.04.1;  % (ISM:) dust,extinction.               
               08.19.4)} % Stars: supernovae, general.
\title{Highlights of the Rome Workshop on Gamma-Ray Bursts
in the Afterglow Era}

\author{D. Q. Lamb}
\institute{Department of Astronomy and Astrophysics, University
of Chicago, \\ 5640 South Ellis Avenue, Chicago, IL 60637}

\date{Received December 15, 1998}

\maketitle

\begin{abstract} 
I review some of the highlights of the Rome Workshop on Gamma-Ray
Bursts, and discuss some of the questions these results pose about the
nature and origin of gamma-ray bursts. 
\end{abstract}

\section{Introduction}

As a result of the initiative and the effort of many people on the
BeppoSAX team, we now live in the era of gamma-ray burst (GRB)
afterglows.  Having advocated for the High Energy Transient Explorer
mission over the years, I know how unlikely it seemed to most
astronomers that GRBs would produce detectable X-ray, let alone
optical, afterglows.  The mounting evidence from the Burst and
Transient Source Experiment onboard the Compton Gamma-Ray Observatory
that GRBs are extragalactic only seemed to strengthen this view.  The
subsequent discoveries that GRBs have X-ray, optical and radio
afterglows has transformed the field.  This workshop shows how great
the impact is that these discoveries have had on the study of GRBs. 
The vast majority of the observational and theoretical results that
were presented at this Workshop come from, or are motivated by, studies
of the radio, optical and X-ray properties of afterglows and host
galaxies, the latter identified by their positional coincidence with
the afterglows.

Here I describe some of the highlights of the Workshop, and discuss
some of the questions these results pose about the nature and origin of
GRBs.  Of necessity, this review reflects my personal point of view. 
Also, I cannot discuss all of the important observational and
theoretical results reported at this meeting, given the limited space
available.  This summary is regretfully therefore incomplete.
%I hope that what I say will be very entertaining and only mildly
%provocative; but fear it may be only mildly entertaining and very
%provocative. 

\section{How Many Classes of GRBs Are There?}

The discovery six months ago of an unusual Type Ic supernova, SN 1998bw
(Galama et al. 1998, Galama 1999), in the BeppoSAX WFC error circle for
GRB 980425 (Soffitta et al. 1998) (see Figure 1) has focused attention
once again on the question: How many distinct classes of GRBs are
there?.  

If GRB 980425 were associated with SN 1998bw, the luminosity of the
burst would be $\sim 10^{46}$ erg s$^{-1}$ and its energy would be
$\sim 10^{47}$ erg.  These values are five orders of magnitude less
than those of the other BeppoSAX bursts, whose luminosities range from
$10^{50}$ to $10^{53}$ ergs s$^{-1}$ and whose energies range from
$10^{52}$ to $10^{55}$ ergs (see below).  Moreover, the behaviors of
the X-ray and optical afterglows would be very different from those of
the other BeppoSAX bursts, yet the burst itself is indistinguishable
from other BeppoSAX and BATSE GRBs with respect to duration, time
history, spectral shape, peak flux, and fluence (Galama et al. 1998).

There is another troubling aspect about the proposed association
between GRB 980425 and SN 1998bw:  Also inside the BeppoSAX WFC error
circle was a fading X-ray source (Pian et al. 1998a,b; Pian et al.
1999; Piro et al. 1998)(see Figure 1).  Connecting this fading X-ray
source with the burst gives a power-law index of $\sim 1.2$ for the 
temporal decay rate (Pian et al. 1998b), which is similar to the
behavior of the other X-ray afterglows observed using BeppoSAX, ROSAT
and ASCA.  This fading X-ray source must therefore be viewed as a
strong candidate for the X-ray afterglow of GRB 980425.  There is also
strong statistical evidence that all Type Ib-Ic supernovae (SNe) do not
produce observable GRBs (Graziani, Lamb \& Marion 1999a,b).

\begin{figure}
\resizebox{\hsize}{!}{\includegraphics{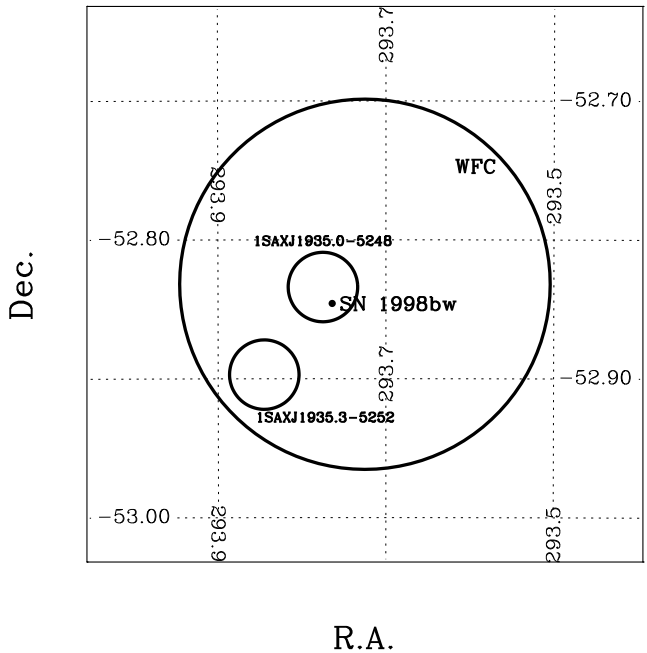}}
\caption
{GRB 980425 field, showing the positional error circle for GRB 980425
determined using the BeppoSAX WFC (large solid circle), the positional
error circles for the fading X-ray source detected by the BeppoSAX NFI
(small solid circle labeled 1SAX1935.3-5252) and for the host galaxy of
SN 1998bw (small solid circle labeled 1SAX1935.0-5248).  From Graziani,
Lamb \& Marion (1999a).
\label{grb980425fig}}
\end{figure}

Approaching the possible association between SN 1998bw and GRBs from
the opposite direction, one can ask:  What fraction $f_{\rm GRB}$ of
the GRBs detected by BATSE could have been produced by Type Ib-Ic SNe,
assuming that the proposed association between GRB 980425  and SN
1998bw is correct, and that the bursts produced are similar to GRB
980425?  Assuming that the association between SN1998bw and GRB 980425
is real, using this association to estimate the BATSE sampling  distance
for such events under the admittedly dubious assumption that the GRBs
produced by Type Ib-Ic SNe are roughly standard candles, and assuming
that {\it all} Type Ib-Ic SNe produce observable GRBs,  Graziani, Lamb
\&( Marion 1999a,b) find that no more than $\sim 90$ such events could
have been detected by BATSE during the lifetime of the {\it Compton}
Gamma-Ray Observatory, indicating that the fraction $f_{\rm GRB}$ of
such events in the BATSE catalog can be no more than about 5\%.  This
result suggests that the observation of another burst like GRB 980425
is unlikely to happen any time soon, even assuming that the association
is real, and  consequently, the question of whether Type Ib-Ic SNe can
produce extremely faint GRBs is likely to remain open for a long time.

Earlier studies have shown that gamma-ray bursts can be separated into
two classes: short, harder, more variable bursts; and long, softer,
smoother bursts (see, e.g., Lamb, Graziani \& Smith 1993; Kouveliotou 
et al. 1993).  Recently, Mukherjee et al. (1999) have provided evidence
for the possible existence of a third class of bursts, based on these
same properties of duration, hardness and smoothness properties of the
bursts.  Also, the hardest long bursts exhibit a pronounced deviation
from the -3/2 power-law required for a homogeneous spatial distribution
of sources, whereas the short bursts and the softest long bursts do not
(Pizzichini 1995; Kouveliotou 1996; Belli 1997, 1999; Tavani 1998). 
These results contradict the expectation that the most distant bursts
should be the most affected by cosmological energy redshift and time
dilation.  Some bursts show considerable high-energy ($E > 300$ keV)
emission whereas others do not, but it is doubtful that this difference
signifies two separate GRB classes, since a similar difference in
behavior is seen for peaks within a burst (Pendleton et al. 1998).

It is not clear whether the short and long classes, and the other
differences among various burst properties, reflect distinct burst
mechanisms, or whether they are due to beaming--or some other property
of the bursts--and different viewing angles.  Some theorists say,
however, that the ``collapsar'' or ``hypernova'' model cannot explain
the short bursts (see, e.g., Woosley 1999).

Because of observational selection effects, all of the GRBs that have
been detected by the BeppoSAX GRBM and observed by the WFC have been
long bursts.  It may be possible for BeppoSAX to revise its GRB
detection algorithm in order to detect short bursts.  We also expect
that HETE-2 will detect short bursts and determine their positions
(Kawai et al. 1999, Ricker et al. 1999).  If so, follow-up observations
may well lead to a breakthrough in our understanding of the nature of
the short bursts similar to that which has occurred for the long
bursts. 

A nightmare I sometimes have is that HETE-2 provides accurate positions
for a number of short bursts, but the positions are not coincident with
any host galaxies because the bursts are due to merging compact object
binaries that have drifted away from their galaxy of origin (see
below).  And furthermore, the bursts exhibit no soft X-ray, optical, or
radio afterglows because any envelope that the progenitors of the
compact objects might have expelled has been left behind, and the
intergalactic medium is too tenuous to dissipate efficiently the energy
in the relativistic external shock that is widely thought to be the
origin of GRB afterglows.  The redshifts of such bursts would be
difficult, if not impossible, to determine, since they could not be
inferred from the redshift of any host galaxy, nor constrained by the
observation of absorption-line systems in the spectrum of any optical
afterglow.

On a more positive note, future radio, optical, and X-ray observations
of GRB afterglows and host galaxies, may well lead to the
identification of new subclasses of GRBs.

\section{GRB Host Galaxies}

\begin{figure}
\resizebox{\hsize}{!}{\includegraphics{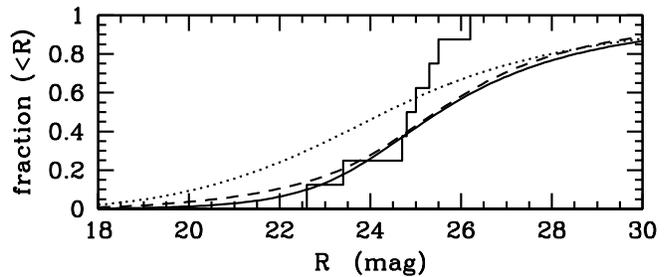}}
\caption
{Cumulative distribution of extinction-corrected R-band magnitudes of
the eight GRB host galaxies identified so far (solid histogram),
compared to that expected for the fiducial star-formation-rate model 
(solid curve), total-stellar-density  model (dotted curve) and
constant-comoving-volume model (dashed curve).  From Hogg \& Fruchter
(1999).
\label{hoggfig}}
\end{figure}

The detection of burst X-ray and optical afterglows has led in eight
cases to identification of the likely host galaxy by positional
coincidence with the optical afterglow.  At $R = 25.5 -26$, the typical
R-band magnitudes of these galaxies, galaxies cover 10-15\% of the sky
for ground-based observations, because of smearing of the galaxy images
due to seeing.  Therefore one expects 1/10 to 1/7 of ground-based
``identifications'' to be incorrect.  If we are lucky, all of the
identifications made to date are correct, but if we are unlucky, one or
two are incorrect.  On the other hand, it is  highly probable that all
of the host galaxies identified from HST observations are correct
(e.g., the host galaxies for GRBs 970228, 970508, 971214, and 980329),
since HST images are free of the effects of seeing that bedevil
observations from the ground.  It is also reassuring that in two cases 
(GRBs 970508 and 971214), the host galaxy identified from ground-based 
observations has been confirmed by later HST observations.

Let me mention a related concern.  Until very recently, all GRB host
galaxies had $R = 25.7 \pm 0.3$, no matter what their redshift and no
matter how the afterglow on which the identification is based was
discovered (i.e., whether detected in the optical, NIR, or radio); that
is, the R-band magnitude of the GRB host galaxy appeared to be a kind
of ``cosmological constant.''  In contrast, if the GRB rate is
proportional to the star formation rate (SFR)(see below), one  expects
a relatively broad distribution of R-band magnitudes for GRB host
galaxies (Hogg \& Fruchter 1999, Fruchter 1999, Madau 1999)(see Figure
2).   The recent discovery of the host galaxy of GRB 980703 at $R =
22.6$ broadens the observed distribution of host galaxy R-band
magnitudes, provided the identification is correct.  However, it also
increases the asymmetry of the R-band magnitude distribution, which
exhibits a tail toward the bright end  and a cutoff toward the dim. 
This is the opposite of what one expects if the GRB rate is
proportional to the SFR.  

This raises the possibility that in some cases we are merely finding
the first galaxy along the line-of-sight to the burst.  If so, in some
cases the galaxy found may be a foreground galaxy, and the actual host
galaxy may lie behind it.  Or it might be that the GRB rate is not 
proportional to the SFR (see the discussion below).  Or most likely of
all, the asymmetry may merely reflect the fact that we are still very
much in the regime of small number statistics.  Additional
confirmations and/or identifications of host galaxies using HST will
resolve this question.

\begin{figure}
\includegraphics[width=0.45\textwidth]{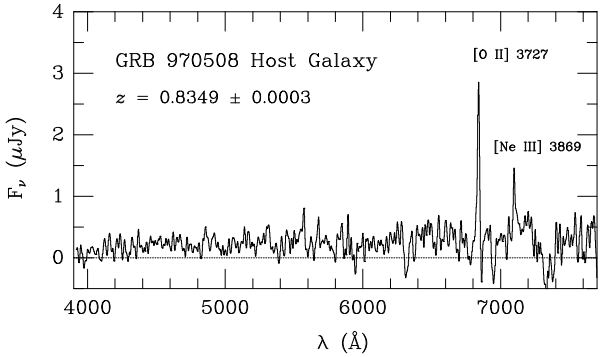}
\caption
{The weighted average spectrum of the host galaxy of GRB 970508. 
Prominent emission lines of [O II] and [Ne III] are labeled.  From 
Bloom et al. (1998).
\label{grb970508_emission.ps}}
\end{figure}

Castander and Lamb (1998) showed that the light from the host galaxy of
GRB 970228, the first burst for which X-ray and optical afterglows were
detected, is very blue, implying that the host galaxy is undergoing
copious star formation and suggesting an association between GRB
sources and star forming galaxies.  Subsequent analyses of the color of
this galaxy (Castander \& Lamb 1999; Fruchter et al. 1999; Lamb, 
Castander \& Reichart 1999) and other host galaxies (see, e.g, Kulkarni
et al. 1998; Fruchter 1999) have strengthened this conclusion, as does
the morphology and the detection of [OII] and Ly${\alpha}$ emission
lines from several host galaxies (see, eg., Metzger et al. 1997b;
Kulkarni et al. 1998; Bloom et al. 1998) (see Figure 3).  The
positional coincidences between several burst afterglows and the bright
blue regions of the host galaxies (see Figure 4), and the evidence for
extinction by dust of some burst afterglows (see, e.g., Reichart 1998;
Kulkarni et al. 1998; Lamb, Castander \& Reichart 1999), suggests that
these GRB sources lie near or in the star-forming regions themselves.

\begin{figure}
\resizebox{\hsize}{!}{\includegraphics{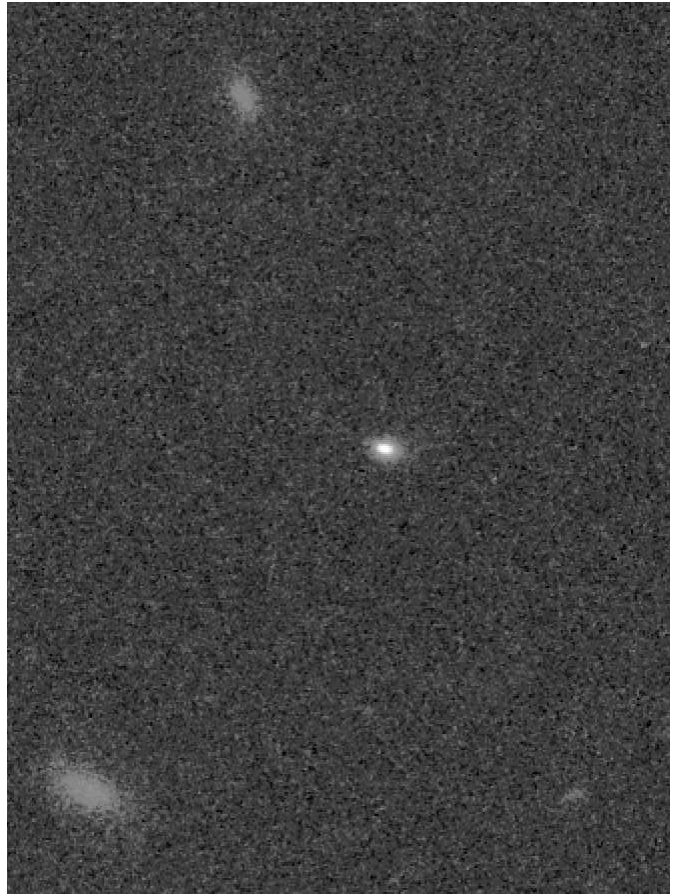}}
\caption
{HST image of the afterglow and host galaxy of GRB 970508 taken 1998 August.  
From Fruchter (1999).
\label{grb970508_stis_aug98_greyscale}}
\end{figure}

The inferred size ($R \lesssim 1 - 3$ kpc) and the morphology of GRB
host galaxies strongly suggest that they are primarily low-mass ($M
\lesssim 0.01 M_{\rm Galaxy}$) but not necessarily sub-luminous
galaxies, because of the ongoing star formation in them (most have $L
\lesssim 0.01 - 0.1 L_{\rm Galaxy}$, but some have $L \sim L_{\rm
Galaxy}$; here $M_{\rm Galaxy}$ and $L_{\rm Galaxy}$ are the mass and
luminosity of a galaxy like the Milky Way).  A point  sometimes not
fully appreciated is that, while the total star formation rate in GRB
host galaxies is often modest (resulting in modest [OII] and
Ly${\alpha}$ emission line strengths), the star formation {\it per unit
mass} in them is very large.

\section{GRB Distance Scale and Rate}

The breakthrough made possible by the discovery that GRBs have X-ray 
(Costa et al. 1997), optical (Galama et al. 1997) and radio (Frail et
al. 1997) afterglows cannot be overstated.  The discovery by Metzger et
al. (1997a) of redshifted absorption lines at $z = 0.83$ in the
optical spectrum of the GRB 970508 afterglow showed that most, perhaps
all, GRB sources lie at cosmological distances.  Yet we must remember
that GRB 970508 remains the only GRB whose distance we have measured
directly.  The current situation is summarized below, in order of
increasing uncertainty in the redshift determination.

In the cases of two other bursts, GRB 980703 (Bloom et al. 1999) and
GRB 971214 (Kulkarni et al. 1998, Kulkarni 1999), we infer the
redshifts ($z = 0.96$ and 3.42)  of the bursts from the redshift of a
galaxy coincident with the burst afterglow (and therefore likely to be
the host galaxy -- but recall my earlier comments).  

In the case of a fourth burst, GRB 980329, a redshift $z \approx 5$ 
was inferred by attributing the precipitous drop in the flux of the
optical afterglow between the I- and R-bands to the Ly$\alpha$ forest
(Fruchter 1999; Lamb, Castander \& Reichart 1999).  However, Djorgovski
et al. (1999) recently reported that this burst must lie at a redshift
$z < 3.9$, based on the absence of any break longward of 6000 \AA\ in
the spectrum of the host galaxy.

In the case of GRB 980703 (Piro 1999) and of a fifth burst, GRB 980828
(Yoshida 1999), there are hints of an emission-like feature in the
X-ray spectrum, which, if interpreted as a redshifted Fe K-shell
emission line, would provide redshift distances for these bursts. 
However, substantial caution is in order because the statistical
significance of these features is slight.  Indeed, all three
``indirect'' means of establishing the redshift distances of GRBs need
verification by cross-checking the redshift distances found using these
methods against those measured directly using redshifted absorption
lines in the optical spectra of their afterglows.  One arcminute or
better angular positions in near real-time, like those that HETE-2 will
provide (Ricker et al. 1999, Kawai et al. 1999), will greatly
facilitate this task.  

The table below summarizes the current situation, in order of
increasing uncertainty in the redshift determination:

\[
\begin{array}{p{0.85\linewidth}r}
\leftskip 10pt Redshifts of Afterglows & 1 \\
\leftskip 10pt Redshifts of Coincident Galaxies & 2 \\
\leftskip 10pt Redshifts from Afterglow Broad-Band Spectra (?) & 0 \\
\leftskip 10pt Redshifts from Fe Lines in X-Ray Afterglows (??) & 0 \\
& $----$ \\
\leftskip 10pt Total & 3
\end{array}
\]

Even with the paucity of GRB redshift distances currently known, and
the uncertainties about these distances, it is striking how our
estimate of the GRB distance scale continues to increase.  Not so long
ago, adherents of the cosmological hypothesis for GRBs favored a
redshift range $0.1 \lesssim z \lesssim 1$, derived primarily from the
brightness distribution of the bursts under the assumption that GRBs
are standard candles.  (Of course, adherents of the galactic hypothesis
argued for much smaller redshifts!).  Now we routinely talk about
redshift distances in the range $2 \lesssim z \lesssim 6$, and such a
redshift range is supported by the three burst redshifts that have
been determined so far.

Much of the motivation for considering such a redshift range for GRBs
comes from the appealing hypothesis that the GRB rate is proportional
to the star-formation rate (SFR) in the universe, an hypothesis that
arose partly in response to the accumulating evidence, described
earlier, that GRBs occur in star-forming galaxies, and possibly near or
in the  star-forming regions themselves.  How far have we been able to
go in testing this hypothesis?  The answer:  Not very far.  First of
all, as Madau (1999) discussed at this meeting, our knowledge of the
SFR as a function of redshift is itself as yet poorly known.  The few
points derived from the relatively small Hubble Deep Field may not be
characteristic of the SFR in the universe at large, not to mention
concerns about star-forming galaxies at high redshift whose light might
be extinguished by dust in the star-forming galaxies themselves, as
well as uncertainties in the epoch and magnitude of star formation in
elliptical galaxies.  Second, we have redshift determinations for only
four GRBs and R-band magnitudes for only eight GRBs.  Much further work
establishing the star formation rate as a function of redshift in the
universe, as well as the redshift distances for many more GRBs, will be
needed before this hypothesis can really be tested.

One thing is now clear:  GRBs are a powerful probe of the high-$z$
universe.  GRB 971214 would still be detected by BATSE and would be
detected by HETE-2 at a redshift distance $z \approx 10$, and it would
be detected by Swift (whose sensitivity threshold is a factor of 5
below that of BATSE and HETE-2) at $z \approx 20$!  If GRBs are
produced by the collapse of massive stars in binaries, one expects them
to occur out to redshifts of at least $z \approx 10 - 12$, the
redshifts at which the first massive stars are thought to have formed,
which are far larger than the redshifts expected for the most distant
quasars.  The occurrence of GRBs at these redshifts may give us our
first information about the earliest generation of stars; the
distribution of absorption-line systems in the spectra of their {\it
infrared} afterglow spectra will give us information about both the
growth of metallicity at early epochs and the large-scale structure of
the universe, and the presence or absence of the Lyman-$\alpha$ forest
in the {\it infrared} afterglow spectra will place constraints on the
Gunn-Peterson effect and may give us  information about the epoch at
which the universe was re-ionized (Lamb \& Reichart 1999a).

The increase in the GRB distance scale also implies that the GRB
phenomenon is much rarer than was previously thought.  This
implication has been noted at this meeting by Schmidt (1999), who finds
that the GRB rate  must be
\begin{equation}
R_{\rm GRB} \sim 10^{-11} {\rm GRBs}\ {\rm yr}^{-1}\ {\rm Mpc}^{-3}
\end{equation}
in order both to match the brightness distribution of the bursts and to
accommodate the redshift distance of $z = 3.42$ inferred for GRB 971214.

By comparison, the rates of neutron star-neutron star (NS-NS) binary
mergers (Totani 1999) and the rate of Type Ib-Ic supernovae (Cappellaro
et al. 1997) are
\begin{equation}
R_{\rm NS-NS} \sim 10^{-6}\ {\rm mergers}\ {\rm yr}^{-1}\ {\rm Mpc}^{-3}
\end{equation}
\begin{equation}
R_{\rm Type\ Ib-Ic} \sim 3 \times 10^{-5}\ {\rm SNe}\ {\rm yr}^{-1}\ 
{\rm Mpc}^{-3}  \; .
\end{equation}
The rate of neutron star-black hole (NS-BH) binary mergers will be
smaller.  Nevertheless, it is clear that, if either of these events are
the sources of GRBs, only a tiny fraction of them produce an observable
GRB.  Even if one posits strong beaming (i.e., $f_{\rm beam} \approx
10^{-2}$; see below), the fraction is small:
\begin{equation}
R_{\rm GRB}/R_{\rm NS-NS} \sim 10^{-3}\ (f_{\rm beam}/10^{-2})^{-1} 
\end{equation}
\begin{equation}
R_{\rm GRB}/R_{\rm Type\ Ib/c} \sim 3 \times 10^{-5}\ 
(f_{\rm beam}/10^{-2})^{-1} \; .
\end{equation}
Therefore, if such events are the sources of GRBs, either beaming must
be incredibly strong ($f_{\rm beam} \sim 10^{-5} - 10^{-3}$) or only
rarely are the physical conditions necessary to produce a GRB
satisfied.  Can any theoretical astrophysicist be expected to explain
such incredible beaming, or alternatively, such a non-robust, ``flaky''
phenomenon?  I have a solution -- at least in the case of SNe:  We
theorists merely need define those supernovae that produce GRBs to be 
a new class of SNe (Type I$_{\rm grb}$ SNe), and then challenge the
observers to go out and find the other observational criteria that
define this class!

\section{Implied Energies and Luminosities}

The maximum energy $({E_{\rm GRB}})_{\rm max}$ that has been observed
for a GRB imposes an important requirement on GRB models, and is 
therefore of great interest to theorists.  $({E_{\rm GRB}})_{\rm max}$
has increased as the number of GRB redshift distances that have been
determined has increased. Currently, the record holder is GRB 971214
at $z=3.4$, which implies  $E_{\rm GRB} \sim 5 \times 10^{53}$ erg from
its gamma-ray fluence, assuming isotropic emission and $\Omega_M = 0.3$
and $\Omega_\Lambda = 0.7$ (Kulkarni 1999).

The table below summarizes the redshifts and energies of the bursts 
for which these are currently known:

\[
\begin{array}{ccc}
\null
\ \ \ \ \ \ \ \ \ \ $Gamma-Ray\ \ Burst\ $ & z\ & $Energy (if isotropic)$ \\
& & \\
\ \ \ \ \ \ \ \ \ \ 970508\ \  & 0.835\ & 7  \times 10^{51}\ {\rm erg} \\
\ \ \ \ \ \ \ \ \ \ 971214\ \  & 3.42 & 5 \times 10^{53}\ {\rm erg} \\
\ \ \ \ \ \ \ \ \ \ 980703\  \ & 0.966\ & 8 \times 10^{52}\ {\rm erg} \\
\end{array}
\]

This kind of energy is difficult to accommodate in NS-NS or NS-BH
binary merger models without invoking strong beaming.  ``Collapsar'' or
``hypernova'' models have an easier time of it, and can perhaps reach
$\sim 10^{54}$ erg without invoking strong beaming by assuming a high
efficiency for the conversion of gravitational binding energy into
gamma-rays (Woosley 1999).

Both classes of models can be ``saved'' by invoking strong beaming
($f_{\rm beam} \sim 1/10 - 1/100$ (but see  the lack of evidence of
beaming discussed below).  Even if GRBs are  strongly beamed, they are
still far and away the brightest electromagnetic  phenomenon in the
Universe, as the following comparison illustrates:

\hfill
\begin{minipage}{6.5cm}
\begin{itemize}
\item[$\bullet$] $L_{\makebox[0.28in]{\rm SNe}} \lesssim 10^{44}\ 
{\rm erg}\ {\rm s}^{-1}$
\item[$\bullet$] $L_{\makebox[0.28in]{\rm SGR}} \lesssim 10^{45}\ 
{\rm erg}\ {\rm s}^{-1}$
\item[$\bullet$] $L_{\makebox[0.28in]{\rm AGN}} \lesssim 10^{45}\ 
{\rm erg}\ {\rm s}^{-1}$
\item[$\bullet$] $L_{\makebox[0.28in]{\rm GRB}} \sim  10^{51}\ 
(f_{\rm beam}/10^{-2})\ {\rm erg}\ {\rm s}^{-1}$
\end{itemize}
\end{minipage}
\hfill
\\

The luminosities of GRB 970508 and GRB 971214 differ by a factor of 
about one hundred.  Thus (if there was previously any doubt),
determination of the redshift distances for just three GRBs has put to
rest once and for all the idea that GRBs are ``standard candles.''  The
extensive studies by Loredo \& Wasserman (1998a,b) and the study by
Schmidt (1999) reported at this workshop show that the luminosity
function for GRBs can be, and almost certainly is, exceedingly broad,
with $\Delta L_{\rm GRB}/L_{\rm GRB} \gtrsim 10^3$.  The results of
Loredo \& Wasserman (1998a,b) show that the burst luminosity function
could be far broader; and indeed, if GRB 980425 is associated with
SN 1998bw, $\Delta L_{\rm GRB}/L_{\rm GRB} \gtrsim 10^5$.

Even taking a luminosity range $\Delta L_{\rm GRB}/L_{\rm GRB} \gtrsim
10^3$ implies that $\Delta F_{\rm GRB}/F_{\rm GRB} \gtrsim 10^4$, given
the range in the distances of the three GRBs whose redshifts are
known.  This is far broader than the range of peak fluxes in the BASTE
GRB sample, and implies that the flux distribution of the bursts
extends well below the BATSE threshold.

The enormous breadth of the luminosity function of GRBs suggests that
the differences (such as time stretching and spectral softening)
between the apparently bright and the apparently dim bursts are due to
{\it intrinsic} differences  between {\it intrinsically} bright and
faint bursts, rather than to cosmology.

Finally, a broad luminosity function is naturally expected in models
with ultra-relativistic radial outflow and strong beaming (jet-like
behavior).  But then why is no large special relativistic Doppler
redshift seen in GRB spectra; i.e., why is the spread in $E_{\rm peak}$
so narrow?

\section{Burst Models}

NS-NS or NS-BH binary mergers and ``collapsar'' or ``hypernova'' events
continue to be the leading models for the energy source of GRBs.  Rees
(1999) described what he termed the ``best buy'' model, which involves
a NS-BH binary merger and a magnetically powered jet.  Woosley (1999)
reported a series of calculations and hydrodynamic simulations that
explore various stages of the collapsar scenario, including the
production of a hydrodynamic jet (although the jet might also be
magnetically powered in this scenario, if magnetic fields were
included).

The increasingly strong evidence that the  bursts detected by BeppoSAX
originate in galaxies undergoing star formation, and may occur near or
in the star-forming regions themselves, favors the collapsar model and
disfavors the binary merger model as the explanation for long, softer,
smoother bursts.  Simulations of the kicks given to NS-NS and NS-BH
binaries by the SNe that form them shows that most binary mergers are
expected to occur well outside any galaxy (Bulik \& Belczynski 1999). 
This is particularly the case, given that the GRB host galaxies
identified so far have small masses, as discussed earlier, and
therefore low escape velocities.  The fact that all of the optical
afterglows of the BeppoSAX bursts are coincident with the disk of the
host galaxy therefore also disfavors the binary merger model as the
explanation for the long, softer, smoother bursts.

%The titles of the papers presented at this workshop show that it is
%apparently very unpopular to work on the GRB mechanism itself.  The
%reason:  It's too difficult!  Consequently, many people are developing
%or testing models of the burst afterglows, but few are doing the same
%for models of the bursts themselves.

Current models of the bursts themselves fall into three general
categories:  Those that invoke a central engine, those that invoke
internal shock waves in a relativistically expanding wind, and those
that invoke a relativistic external shock wave.  Dermer (1999) argued
that the external shock wave model explains many of the observed
properties of the bursts.  By contrast, Fenimore (1999; see also
Fenimore et al. 1999) argued that several features of GRBs, such as the
large gaps seen in burst time histories, cannot be explained by the
external shock wave model, and that the bursts must therefore be due
either to a central engine or to internal shocks in a relativistically
expanding wind.  Either way, the intensity and spectral variations seen
during the burst must originate at a central engine.  This implies that
the lifetime of the central engine must in many cases be $t_{\rm
engine} \gtrsim 100 - 1000$ s, which poses a severe difficulty for
NS-NS or NS-BH binary merger models, if such models are invoked to
explain the long, softer, smoother bursts, and may pose a problem for
the collapsar model.  Fenimore (1999) reported at this meeting that he
finds no evidence of relativistic expansion in the time histories and
spectra of the GRBs themselves, presenting a possible difficulty for
the internal shock wave model.  

One puzzle about the bursts themselves is:  Why are GRB spectra so
smooth? The shock synchrotron model agrees well with observed burst
spectra.  But this agreement is surprising, since strong deviations from
the simplest spectral shape are expected due to inverse Compton
scattering, and if forward and reverse shock contributions to the
prompt gamma-ray emission occur simultaneously or at different times
(Tavani 1999).

Another puzzle is:  Why is the spread in the peak energy $E_{\rm peak}$
of the burst spectra so narrow?  In the external shock model, this
requires that all GRBs have nearly the same ultra-relativistic value of
$\Gamma$.  The narrow range in $E_{\rm peak}$ is, if anything, more
difficult to understand in the internal shock model.  If $\Delta 
\Gamma/\Gamma << 1$ in the relativistic outflow, the range in $E_{\rm
peak}$ will be narrow, but then it is hard to understand why most of
the energy of the relativistic outflow is dissipated during the burst
rather than in the afterglow.  Conversely, if $\Delta \Gamma/\Gamma >>
1$ in the relativistic outflow, most of the energy of the relativistic
outflow is dissipated during the burst rather than in the afterglow,
but then one expects a wide range of $E_{\rm peak}$'s.  This is a hint
-- like the problem discussed earlier that one would expect strong
beaming to produce a large special relativistic Doppler redshift, yet
this is not  seen in burst spectra -- that there may be something
missing in our picture of the dissipation and radiation mechanisms in
GRBs.

\begin{figure}
\resizebox{\hsize}{!}{\includegraphics[trim=55 400 50 55]{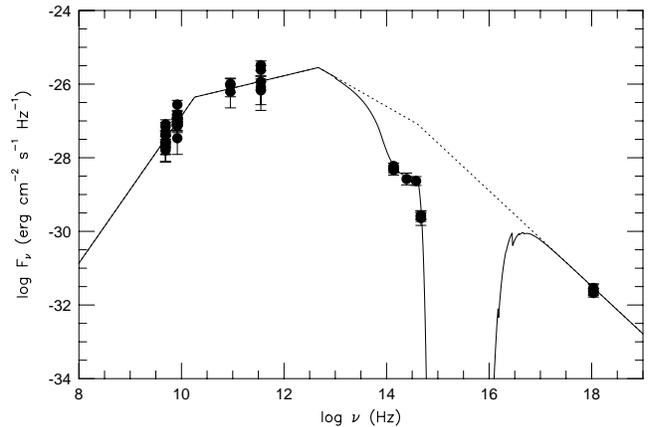}}
\caption
{The radio through X-ray spectrum of the afterglow of GRB 980329.  All
measurements have been scaled to a common time, approximately three
days after the GRB.  The solid curve is the best fit spectrum for an
isotropic fireball that expands into a homogeneous external medium,
extincted by dust at a redshift of $z = 3.5$.  The dotted curve is the
un-extincted spectrum.  From Lamb, Castander \& Reichart (1999).
\label{romefig}}
\end{figure}

\section{Beaming}

Most theorists {\it expect} GRBs to be significantly beamed -- many
energetic astrophysical phenomena are (examples include young
protostars; the so-called ``microquasars,'' which are black hole
binaries in the Galaxy; radio galaxies; and AGN).  And theorists {\it
desire} beaming because it saves their models.  Several speakers at
this workshop have emphasized these points (see, e.g., Dar 1999,
Fargion 1999, and Rees 1999).  Strong beaming probably requires strong
magnetic fields, but no detailed physical model of how this might
happen has been put forward as yet.

One can ask:  Where is the observational evidence for beaming...? 
Fenimore (1999) told us that there is none in the time histories of the
bursts themselves.  Worse yet, Greiner (1999) reported that $f_{\rm
GRB}/f^{\rm X-ray}_{\rm afterglow} \gtrsim 1$ from an analysis of the
ROSAT all sky survey.  This constraint may not be as strong as it
appears, because the duration of the temporal exposure in the ROSAT 
all-sky survey is only a few hundred seconds, and thus the sensitivity
of the survey is relatively poor.  Consequently, soft X-ray afterglows
would be detectable by the ROSAT all-sky survey only within a day or so
after the burst, when (in the relativistic external shock model of
afterglows -- see below) the soft X-ray emission is still highly
beamed.

Constraints on so-called ``orphan'' optical afterglows, and therefore
on the beaming of GRBs, will be strengthened by new low-$z$ SN Ia
searches that will soon be underway.  These searches will monitor an
area of the sky that is roughly ten times larger than that monitored by
current high-$z$ SN Ia searches down to the same limiting magnitude
($m_B \approx 20$) (Perlmutter 1999).

%Given the crucial importance of beaming, I would like to issue an
%appeal to the observers:  Be the theorists' friend... and {\it please
%find evidence of beaming}!

\section{Afterglow Models}

The simple fireball model (i.e., a spherically symmetric relativistic
external shock wave  expanding into a homogeneous medium) has been ``in
and out of the hospital'' for months, but notices of its death appear
to be premature.  This amazes me, given the wealth of complexities one
can easily imagine in the fireball itself and in its environment
(M\'esz\'aros 1999).  If the simple relativistic fireball model (or
even more complex variants of it) suffice to explain burst afterglows
(see Figure 5), much can be learned, including the energy of the
fireball per unit solid angle, the ratio of the energy in the magnetic
field to that in relativistic electrons, and the density of the
external medium into which the fireball expands (Wijers \& Galama 1999;
van Paradijs 1999; Lamb, Castander \& Reichart 1999).  It should be
possible, in principle, to use the effects on the afterglow spectrum of
extinction due to dust in the host galaxy and of absorption by the
Lyman-$\alpha$ forest to determine the redshift of the burst itself,
but so far, this goal has eluded modelers (see, e.g, Lamb, Castander \&
Reichart 1999). 

Currently, we are in the {\it linear} regime in terms of what we learn
from each individual {\it afterglow} because, given the diversity of
GRBs, GRB afterglows, and host galaxies, we have yet to sample the full
``phase space'' of afterglow or host galaxy properties.  Still less
have we sampled the full ``phase space'' of combinations of burst,
afterglow, and host properties.  

At the same time, we are in the strongly {\it non-linear} regime, in
terms of what we learn from each individual {\it observation} of a
burst afterglow.  The value of each astronomer's observation is
enhanced by the observations made by all other astronomers.  As we have
heard from several speakers at this workshop, the amount of information
that can be gleaned from a given afterglow depends greatly on the
number of  measurements that exist both simultaneously in time and in
wavelength, from the radio  through the millimeter, sub-millimeter,
near-infrared, optical, and X-ray bands.  Furthermore, since the range
of redshifts for the bursts (and therefore also their afterglows) is
large, we cannot know in advance which bands will be crucial.  Thus
simultaneous or near-simultaneous multi-wavelength observations of
burst afterglows are essential, and therefore observations  by as many
observers as possible must be encouraged.  Finally, greater
co-operation and co-ordination among observers is important,  and
should be facilitated, as has been done by setting up the invaluable
service represented by the Gamma-Ray Burst Coordinate Network (GCN)
(Barthelmy et al. 1999).

\section{Star-Forming Regions}

Star forming regions consist of a cluster of O/B stars that lie in and
around a clumpy cloud of dust and gas.  We expect $A_V >> 1$ for O/B
stars embedded in the cloud, and $A_V \approx 0$ for O/B stars that
have drifted out of the cloud and/or lie near the surface of the cloud
and have expelled the gas and dust in their vicinity.  Thus the
optical/UV spectrum of star forming regions is a sum of the spectra  of
many hot (blue) stars, some of which are embedded in the cloud, and
therefore heavily extinguished, and some of which lie on the surface or
around the cloud, and are therefore essentially un-extinguished.  This
composite spectrum is rather blue, and yields a value $A_V^{\rm eff}
\approx 1$ when a single extinction curve is fitted to it.

The situation is very different when we consider an individual
line-of-sight, as is appropriate for the afterglow of a GRB.  If the GRB
source lies outside and far away from any star-forming region, we
expect $A_V^{\rm afterglow} \lesssim 1$; if the GRB source lies outside
but near a star-forming region, we expect $A_V^{\rm afterglow} \lesssim
1$ about half the time and $A_V^{\rm afterglow} >> 1$ about half the
time.  Finally, if the GRB source is embedded in the star-forming
region, we expect $A_V^{\rm afterglow} >> 1$. 

Thus, if GRB sources actually lie in star-forming regions, one would
expect $A_V^{\rm afterglow} >> 1$ (values of $A_V \sim 10-30$ are not
uncommon for dense, cool molecular clouds in the Galaxy).  Is this
consistent with what we see?  No.  However, this may not mean that GRB
sources do not lie in star-forming regions.  The reason is that the
soft X rays and the UV radiation from the GRB and its afterglow are
capable, during the burst and immediately afterward, of vaporizing all
of  the dust in their path (Lamb \& Reichart 1999b).  Thus the value of
$A_V^{\rm afterglow}$ that we measure may have nothing to do with the
pre-existing value of the extinction through the star-forming region in
which the burst source is embedded, but may instead reflect merely the
extinction due to dust and gas in the disk of the host galaxy.

The GRB, and its soft X-ray and UV afterglow, are also capable of
ionizing gas in any envelope material expelled by the progenitor of the
burst source and in the interstellar medium of the host galaxy.  This
will produce Str\"omgren spheres or very narrow cones (if the burst and
its afterglow are beamed) in hydrogen, helium and various metals
(Bisnovatyi-Kogan \& Timokhin 1998, Timokhin \& Bisnovatyi-Kogan 1999,
M\'esz\'aros 1999).  Recombination of the ionized hydrogen eventually
produces intense [CII], [CIV], [OVI] and [CIII] emission lines in the
UV, and intense H$\alpha$ and H$\beta$ emission lines in the optical. 
However, the line fluxes may still not be strong enough to be
detectable at the large redshift distances of GRB host galaxies. 
Interaction of the GRB and its soft X-ray afterglow with any envelope 
material expelled by the progenitor of the burst source and with the
surrounding interstellar medium can also produce intense fluorescent
iron line emission (see, e.g., M\'esz\'aros 1999), but it is again
difficult to see how the line flux could be large enough to be
detectable or to explain the hints of a fluorescent iron emission line
in the X-ray afterglows of GRB 980703 (Piro et al. 1999) and GRB 980828
(Yoshida et al. 1999).

\section{Future}

Each person has their own favorite list of future observational needs.  
Here is mine:

\noindent
$\bullet$ 
We need a high rate ($> 100$ GRBs yr$^{-1}$) of bursts with good 
locations, in order to change the sociology of ground-based optical and
radio observations.  This many good GRB positions to follow-up each
year would make it possible to propose and carry out GRB afterglow
monitoring programs at many medium-to-large aperture telescopes.

\noindent
$\bullet$ 
The diversity of GRBs, GRB afterglows, and host galaxies means that we
need a large number ($> 1000$) of good GRB positions in order to be
able to study the correlations between these properties.  This is
important for determining whether or not there are distinct subclasses
of bursts, and more than one burst mechanism.  Any correlations found
will also impose important constraints on burst mechanisms and models.

\noindent
$\bullet$ 
We need many rapid (near real time) one arcminute GRB positions in
order to determine whether or not significant optical emission
accompanies the bursts (Park 1999), and to make it possible to take
spectra of the burst afterglows while the afterglows are still bright
-- and thereby obtain redshifts of the bursts themselves from 
absorption line systems, and if there are bursts at high redshifts,
from the Ly$\alpha$ break.

\noindent
$\bullet$ 
All of the GRBs that BeppoSAX has detected are ``long'' bursts. 
Currently we know nothing about the afterglow properties, the distance
scale, and the hosts (if any) of ``short'' bursts.  Therefore we need
good/quick positions for short bursts, in order to determine these
properties for short bursts in the same way that BeppoSAX has enabled
us to determine these properties for long bursts.

\noindent
$\bullet$ 
Currently, there is a largely unexplored gap in our knowledge of the
X-ray and optical behavior of burst afterglows of $\sim 10^4 - 10^5$
seconds immediately following the bursts, corresponding to the time
needed to bring the BeppoSAX NFIs to bear on a burst.  We need to fill
in this unexplored gap, in order to see if bursts always, often, or
rarely join smoothly onto their X-ray and optical afterglows, and to
explore the geometry and kinematics of GRB afterglows (Sari 1999).

\noindent
$\bullet$ 
We also need to search for variability in the X-ray and optical
afterglows.  Observations of such variability would impose severe
constraints on models, including the widely-discussed relativistic
fireball model of burst afterglows (see, e.g., Fenimore 1999).

\section{Acknowledgement}

The Rome Workshop provided a feast of observational and theoretical
results, and the opportunity to discuss them.  On behalf of all of the
Workshop participants, I would like to thank Enrico Costa, Luigi Piro,
Filippo Fontana, and everyone else who helped to organize this meeting
for bringing all of us together and for providing us with such ``fine
dining.''


\begin{thebibliography}{}

\bibitem[1999]{barthelmy99}
Barthelmy, S.~D., Butterworth, P., Cline, T.~L. \& Gehrels, N. 1999,
these proceedings

\bibitem[1997]{belli97}
Belli, B.~M. 1997, ApJ, 413, 281

\bibitem[1999]{belli99}
Belli, B.~M. 1999, these proceedings

\bibitem[1997]{bisnovatyi-kogan97}
Bisnovatyi-Kogan, G.~S. \& Timokhin, A.~N. 1997, Astron. Reports, 41,
483

\bibitem[1998]{bloom98}
Bloom, J. et al. 1998, ApJ, 507, L25

\bibitem[1999]{bloom99}
Bloom, J. et al. 1999, these proceedings

\bibitem[1999]{bulik99}
Bulik, T. \& Belczynski, K. 1999, these proceedings

\bibitem[1997]{cappellaro97}
Cappellaro, E. et al. 1997, A\&A, 322, 431

\bibitem[Castander \& Lamb 1998]{CL98} 
Castander, F.~J. \& Lamb, D.~Q. 1998, in Gamma-Ray Bursts, eds. C.~A.
Meegan, R.~D. Preece, and T.~M. Koshut (New York: AIP) 520

\bibitem[Castander \& Lamb 1999]{CL99} 
Castander, F.~J. \& Lamb, D.~Q. 1999, ApJ, in press

\bibitem[1997]{costa97}
Costa, E., et al. 1997, IAU Circular No. 6576

\bibitem[1999]{dar99}
Dar, A. 1999, these proceedings

\bibitem[1999]{dermer99}
Dermer, C. 1999, these proceedings

\bibitem[1999]{djorgovski99}
Djorgovski, S.~G., et al. 1999, invited talk at the Santa Barbara 
Workshop on ``Gamma-Ray Bursts and their Afterglows''

\bibitem[1999]{fargion99}
Fargion, D.  1999, these proceedings

\bibitem[1999]{fenimore99a}
Fenimore, E.~E. 1999, these proceedings

\bibitem[1999]{fenimore99b}
Fenimore, E.~E., et al. 1999, ApJ, in press (astro-ph/9802200)

\bibitem[1997]{frail97}
Frail, D.~A. et al. 1997, Nature, 389, 261

\bibitem[1999]{frail99}
Frail, D.~A. 1999, these proceedings

\bibitem[1999]{fruchter99a}
Fruchter, A.~S. 1999, these proceedings

\bibitem[1999]{fruchter99b}
Fruchter, A.~S., et al. 1999, ApJ, in press (astro-ph/9807295)

\bibitem[1999]{galama99}
Galama, T.~J. 1999, these proceedings

\bibitem[1997]{galama97}
Galama, T.~J., et al. 1997, IAU Circular No. 6584 

\bibitem[1998]{galama98}
Galama, T.~J., et al. 1998, Nature, in press (astro-ph/9806175)

\bibitem[1999a]{graziani99a}
Graziani,~C, Lamb,~D.~Q., and Marion,~G.~H. 1999a, ApJ, submitted
(astro-ph/9810374)

\bibitem[1999b]{graziani99b}
Graziani,~C, Lamb,~D.~Q., and Marion,~G.~H. 1999b, these proceedings

\bibitem[1999]{greiner99}
Greiner, J. 1999, these proceedings

\bibitem[1999]{hogg99}
Hogg, D.~W. \& Fruchter, A.~S. 1999, ApJ, in press (astro-ph/9807262)

\bibitem[1999]{kawai99}
Kawai, N., et al. 1999, these proceedings

\bibitem[1993]{kouveliotou93}
Kouveliotou, C., et al. 1993, ApJ, 413, L101

\bibitem[1996]{kouveliotou96}
Kouveliotou, C., et al. 1996, AIP Conf. Proc. No. 384, 42

\bibitem[1999]{kulkarni99}
Kulkarni, S. 1999, these proceedings

\bibitem[1998]{kulkarni98}
Kulkarni, S. et al. 1998, Nature, 393, 35

\bibitem[1993]{lamb93}
Lamb, D.~Q., Graziani, C. \& Smith, I.~A. 1993, ApJ, 413, L11

\bibitem[1999]{lamb99a}
Lamb, D.~Q., Castander, F.~J. \& Reichart D.~E. 1999, these proceedings

\bibitem[1999]{lamb99b}
Lamb, D.~Q. \& Reichart, D.~E. 1999a, submitted to ApJ

\bibitem[1999]{lamb99c}
Lamb, D.~Q. \& Reichart, D.~E. 1999b, submitted to ApJ

\bibitem[19998a]{loredo98a}
Loredo, T.~J. \& Wasserman, I. 1998a, ApJ, 502, 75

\bibitem[19998b]{loredo98b}
Loredo, T.~J. \& Wasserman, I. 1998b, ApJ, 502, 108

\bibitem[1999]{madau99}
Madau, P. 1999, these proceedings

\bibitem[1997a]{metzger97a}
Metzger, M., et al. 1997a, IAU Circular No. 6655

\bibitem[1997b]{metzger97b}
Metzger, M., et al. 1997b, Nature, 387, 878

%\bibitem[1998]{meegan98}
%Meegan, C. A. et al. 1998, in Gamma-Ray Bursts, AIP Conference Proceedings
%428, eds. C. A. Meegan, R. D. Preece, \& T. M. Kogut (New York: AIP), 3

\bibitem[1999]{meszaros99}
M\'esz\'aros, P. 1999, these proceedings

\bibitem[1999]{mukherjee99}
Mukherjee, S., et al. 1999, ApJ, in press (astro-ph/9802085)

\bibitem[1999]{vanparadijs99}
van Paradijs, J. 1999, these proceedings

\bibitem[1999]{park99}
Park, H.~S. 1999, these proceedings

\bibitem[1998]{pendleton98}
Pendleton, G., et al. 1998, ApJ, in press

\bibitem[1999]{perlmutter99}
Perlmutter, S. 1999, private communication

\bibitem[1998a]{pian98a}
Pian, E., et al. 1998a, GCN Report No. 61

\bibitem[1998b]{pian98b}
Pian, E., Frontera, F., Antonelli, L.~A. \& Piro, L. 1998b, GCN Report 
No. 69

\bibitem[1999]{pian99}
Pian, E., et al. 1999, these proceedings

\bibitem[1998]{piro98}
Piro,~L., et al. 1998, GCN Report No. 155

\bibitem[1999]{piro99}
Piro,~L., et al. 1999, these proceedings

\bibitem[1995]{pizzichini95}
Pizzichini, G. 1995, in Proc. XXIV ICRC 81

\bibitem[1999]{rees99}
Rees, M.~J. 1999, these proceedings

\bibitem[1997]{reichart99}
Reichart, D. E. 1998, ApJ, 495, L99

\bibitem[1999]{ricker99}
Ricker, G., et al. 1999, these proceedings

\bibitem[1999]{sari99}
Sari, R. 1999, these proceedings

\bibitem[Schmidt 1999]{schmidt99}
Schmidt, M. 1999, these proceedings

\bibitem[1998]{soffitta98}
Soffitta, P. et al. 1998, IAU Circ. No. 6884

\bibitem[1998]{tavani98}
Tavani, M. 1998, ApJ, 497, L21

\bibitem[1999]{tavani99}
Tavani, M. 1999, these proceedings

\bibitem[1999]{totani99}
Totani, T. 1999, ApJ, 511, 41

\bibitem[1999]{bisnovatyi-kogan99}
Timokhin, A.~N. \& Bisnovatyi-Kogan, G.~S. 1999, these proceedings

%\bibitem[1998a]{wang98a}
%Wang, L. \& Wheeler, J.~C. 1998a, astro-ph/9806212

%\bibitem[1998b]{wang98b}
%Wang, L. \& Wheeler, J.~C. 1998b, ApJ, 504, L87

\bibitem[1999]{wijers99}
Wijers, R. \& Galama, T.~J. 1999, ApJ, in press

\bibitem[1999]{woosley99}
Woosley, S.~E. 1999, these proceedings

\bibitem[1999]{yoshida99}
Yoshida, A., et al. 1999, these proceedings

\end{thebibliography}
\end{document}